\documentclass[aps,pra,reprint,groupedaddress,amsmath,amssymb,preprintnumbers,showpacs,longbibliography]{revtex4-2}
\usepackage{graphicx}
\usepackage{bm}
\usepackage{dcolumn}
\newcommand{\eq}[1]{Eq.~(\ref{#1})}

\newcommand{\be}{\begin{equation}}
\newcommand{\ee}{\end{equation}}
\newcommand{\ba}{\begin{eqnarray}}
\newcommand{\ea}{\end{eqnarray}}
\newcommand{\bs}{\begin{subequations}}
\newcommand{\es}{\end{subequations}}
\newcommand{\bw}{\begin{widetext}}
\newcommand{\ew}{\end{widetext}}

\usepackage[final]{showlabels}

\begin{document}

\title{Strong-field approximation for high-harmonic generation in infrared laser pulses in the accelerated Kramers-Henneberger frame}

\author{Lars Bojer Madsen}
\affiliation{Department of Physics and Astronomy, Aarhus
University, DK-8000 Aarhus C, Denmark}

\date{\today}

\begin{abstract}
The strong-field approximation for high-harmonic generation in near-infrared and infrared laser pulses is formulated in the accelerated Kramers-Henneberger frame. The accompanying physical picture is discussed and the  nature of the leading-order term is contrasted with that of the three-step model following the strong-field-approximation formulation in the length or velocity gauges. The theory is illustrated by high-harmonic generation spectra for atomic hydrogen. 
\end{abstract}

\maketitle

\section{\label{sec:introduction} Introduction}
When an intense near-infrared or infrared laser pulse interacts with an atomic or molecular gas or with a solid-state sample, the nonlinear laser-matter interaction may lead to the emission of  coherent radiation with a frequency that equals an integer $N\gg 1$ times the fundamental frequency of the driving field. For linearly polarized laser pulses and atomic or molecular targets  some aspects of this high-harmonic generation (HHG) process can be rationalized in terms of the three-step model~\cite{Schafer1993,Corkum1993,Kulander1993}. In the first step of this model, the target is strong-field ionized in a process that can often be envisioned as a tunneling-like process. In the second step, the freed electron propagates in the presence of the laser pulse. The electron may be steered back to the parent ion due to the alternating field direction of the laser pulse. In the third step, the electron recombines and emits its accumulated energy as HHG radiation. The quantum mechanical framework identifying these steps is known as the Lewenstein model or the strong-field approximation (SFA) of HHG~\cite{Lewenstein1994}. In the context of HHG in solids, certain characteristics of the harmonic spectra can be captured by an independent-electron bandstructure model involving (i) interband transitions from the valence to the conduction band, (ii) intraband propagation of electrons in the conduction band and holes in the valence band and (iii) interband recombination into the valence band. Harmonics are emitted as a consequence of inter- and intraband dynamics and the interplay between these dynamics~\cite{Vampa2014}. Both for the atoms and solids, the theory development in Refs.~\cite{Lewenstein1994,Vampa2014} is based on the length gauge (LG) expression for the interaction between the electron and the electric field, $\bm E(t)$, of the external laser pulse, i.e., $V_L^\text{LG}(t)= -q \bm E(t) \cdot \bm r$ with $\bm r$ the electron coordinate and $q=- |e|$ its charge. Of course the physical pictures that emerge from an analysis of these theories depend on the choice of representation of the interaction. For example, the notion of tunneling ionization is natural when $V_L^\text{LG}(t)$ is added to the atomic or molecular potential to form an effective potential with a barrier through which the electron can tunnel. If, on the other hand, the  velocity gauge (VG) form of the laser-electron interaction,  $V_L^\text{VG}(t) = -\frac{q}{m}\bm A(t) \cdot \bm p + \frac{q^2 \bm A(t)^2}{2 m}$, with $\bm A(t)$ the vector potential of the laser pulse, had been considered, the initial ionization step in the three-step model would have been less clearly identified as a tunneling step since no effective spatial tunneling barrier would emerge when adding $V_L^\text{VG}(t)$ to the atomic or molecular potential. Nevertheless, the picture of HHG as a process involving strong-field induced transitions out of the the initial state, electron propagation and recombination emerges naturally in SFA approaches using both gauges. 

The insights extracted from HHG spectra regarding properties of the target often rely on an interpretation in terms of the three-step model. For example in orbital tomography, HHG spectra are used to reconstruct the Dyson orbital of the ionizing system by analyzing the spectra in terms of the recombination matrix element~\cite{Itatani2004}. In HHG spectroscopy, analysis of the spectra based on a many-electron SFA for HHG, which embodies the quantum mechanical three-step picture, may be used to extract information about charge migration in molecules~\cite{Kraus2015}. In the case of solids, ideas analogous to those of the three-step model were instrumental in suggesting an all-optical reconstruction of crystal band structure by HHG~\cite{Vampa2015}. These successes of the three-step SFA model rely on its ability to capture qualitative aspects of the HHG spectra. Quantitatively, its predictions deviate from the results following simulations based on the time-dependent Schr\"odinger equation (TDSE). In particular the accuracy of the SFA is challenged at harmonic orders lower than those close to the cut-off region~\cite{Perez-Hernandez2009}; the former spectral region will be in focus in this work. 

It is in the context of considering the interplay between the form of the interaction, e.g., $V_L^\text{LG}(t)$ or $V_L^\text{VG}(t)$, and the resulting physical model, that it may be worthwhile to consider if alternative physical pictures and associated insights arise for the interpretation of the HHG process if the fundamental electron-laser interaction is expressed in yet a different manner. One such possibility, which is explored in this work,  is the form of the interaction in the accelerated Kramers-Henneberger (KH) frame~\cite{Henneberger1968}, where a unitary operator is used to displace the electron coordinate by the instantaneous position of a free electron in the oscillating external field. The KH frame was used some time ago to interpret HHG spectra but only in the high-intensity, high-frequency stabilization regime~\cite{Reed1993}.  Interestingly, there appears to be no discussion in the literature of the theory of HHG formulated in the KH frame in low frequency infrared and near-infrared fields. It is the purpose of the present work to provide such a discussion and highlight the accompanying physical picture, i.e., to formulate the SFA for HHG in infrared or near-infrared fields in the KH frame. Of particular relevance for this purpose is a very recent work~\cite{Lakhotia2020} which used an analysis related to the leading-order KH frame HHG contribution identified here, to reconstruct valence electron densities and effective potentials from HHG spectra in crystalline magnesium fluoride and calcium fluoride obtained with near-infrared fields. Here a theoretical foundation for such analysis is discussed  including a formal identification of higher-order contributions to the spectra.

The paper is organized as follows. In Sec.~\ref{sec:theory},  the SFA theory for HHG in the KH frame is formulated, discussed, and illustrated by a simple example. Section \ref{Conclusion} concludes and provides an outlook.

\section{\label{sec:theory} Theory}

\subsection{Expressions for HHG spectra and dipole acceleration}
\label{expressions}
For a sufficiently thin target, propagation effects can be neglected~\cite{Gaarde_review} and the HHG spectrum, i.e., the signal strength $S(\omega)$ as a function of harmonic frequency $\omega$,  can be modelled by~\cite{Baggesen2011}
\be
S(\omega) \propto \frac{1}{\omega^2} | \bm{\epsilon} \cdot \ddot{\tilde{\bm D}}(\omega) |^2,
\label{signal}
\ee
where $\bm \epsilon$ describes the polarization component of interest of the generated light and where $\ddot{\tilde{\bm D}}(\omega)$ is given by the Fourier transform of the dipole acceleration
\be
\ddot{\tilde{\bm D}}(\omega) = \frac{1}{2\pi} \int_{-\infty}^{\infty} dt e^{-i\omega t} \frac{d^2}{dt^2} \langle  \bm D(t) \rangle.
\label{D}
\ee
Here $\langle \rangle$ denotes the expectation value of the operator and involves the integral over the degrees of freedom of the quantum state describing the system. In this paper, the tradition of strong-field physics in terms of a description by a single-active electron model with an effective single-electron potential $V(\bm r)$ will be followed. Note that this approach includes the independent-electron band structure theory of solids. In the case under consideration, the dipole operator reads
\be
\bm D = q  \bm r,
\ee
with $q=-|e|$ for an electron. The investigation of $\langle \bm D (t) \rangle$  in an SFA setting was the starting point for the quantum mechanical treatment of the  three-step model of HHG~\cite{Lewenstein1994}. Here,  Ehrenfest's theorem is used to connect the acceleration to the force
\be
\frac{d^2}{dt^2} \langle  \bm D(t) \rangle = q  \langle - \bm \nabla_{\bm r} V(\bm r) \rangle .
\label{dipole}
\ee
At this point, the reader is reminded about some basic relations between different forms of the fundamental laser-matter interaction
 (see, e.g., Ref.~\cite{Madsen2021a} for a recent discussion). 
 A natural starting point is to introduce the laser-matter interaction through the minimal coupling by substituting the canonical momentum $\bm p$ as $\bm p  \rightarrow \bm p - q \bm A(t)$, with $\bm A(t)$ the vector potential of the laser pulse. The electric dipole approximation is often sufficiently accurate to capture dominant effects and hence $\bm A(t)$ does not depend on the spatial coordinate but only on time $t$ as indicated in the notation. The VG Hamiltonian then reads
$H^{\text{VG}}(t) = \frac{(\bm p - q \bm A(t))^2}{2m} + V(\bm r)$ and the TDSE reads 
$i\hbar  \frac{d}{dt}| \psi^\text{VG}(t) \rangle = H^\text{VG}(t) | \psi^\text{VG} (t) \rangle$.
The transformation $| \psi(t) \rangle = T(t) | \psi^\text{VG}(t)\rangle$ into the accelerated KH frame~\cite{Henneberger1968} is performed by the unitary operator  
\be
T (t)= 
e^{\frac{i}{\hbar} \frac{q}{m}  \bm \alpha (t) \cdot \bm p } e^{ \frac{i}{\hbar}\int^t d \tau \frac{q^2 \bm A(\tau)^2}{2m}},
\label{T}
\ee
where $\bm \alpha(t)$ is the spatial quiver amplitude 
\be
\bm \alpha(t) = - \int^t d \tau \bm A(\tau).
\ee
The KH Hamiltonian reads
\be
H(t) = \frac{\bm p^2}{2m} + V(\bm r + \frac{q}{m} \bm \alpha(t)).
\label{H}
\ee
 Symbols without superscript denote quantities in the KH frame. Hence, the symbol $|\psi(t) \rangle$ denotes the solution to the TDSE with the Hamiltonian from \eq{H}.

The operator $T(t)^\dagger T(t) = 1$ is inserted on the right-hand side of \eq{dipole} and effectuates the transformation from the VG to the KH frame as follows $\langle  - \bm \nabla_{\bm r} V(\bm r, t) \rangle  = \langle  T(t)^\dagger \{ T(t) [ - \bm \nabla_{\bm r} V(\bm r) ]T (t)^\dagger \} T(t)  \rangle $ to obtain
\be
\frac{d^2}{dt^2} \langle  \bm D  \rangle = q  \langle \psi(t) | - \nabla_{\bm r}  V(\bm r + \frac{q}{m} \bm \alpha(t)) | \psi(t)  \rangle,
\label{x-space}
\ee
where the bra and ket in the KH frame are explicitly included on the right-hand side.  Of course the result in \eq{x-space} could be stated directly from the Hamiltonian in \eq{H} by  applying Ehrenfest's theorem. The above unitary-transformation perspective might be useful to some readers and is therefore included.

To proceed it is useful to consider the formulation in wavevector space. Here the Fourier transforms between real- and wavevector space, i.e., $\bm r$- and $\bm k$-space are defined with the normalizations  $V(\bm r) = (2 \pi)^{-3/2} \int d \bm k e ^{i \bm k \cdot \bm r} \tilde{V }(\bm k)$ and $\tilde{V}(\bm k) = (2 \pi)^{-3/2}\int d \bm r e^{-i \bm k \cdot \bm r} V(\bm r)$. These relations are used to obtain 
\be
- \nabla_{\bm r}  V(\bm r + \frac{q}{m} \bm \alpha(t))= \frac{-i }{(2 \pi)^{3/2}} \int d \bm k \,   \bm k \tilde{V}(\bm k) e^{i \bm k \cdot (\bm r + \frac{q}{m} \bm \alpha(t))}.
\ee
In $\bm k$-space the expression in \eq{x-space} therefore reads
\be
\frac{d^2}{dt^2} \langle  \bm D  \rangle = -i q \int  d \bm k  \bm k \tilde{V}(\bm k)  e^{i \frac{q}{m} \bm k \cdot \bm \alpha(t) }F(\bm k, t), 
\label{dipole-KH}
\ee
where a time-dependent generalization of the (inelastic) form factor is identified by
\be
F(\bm k, t) = \langle \psi(t) | \frac{e^{i \bm k \cdot \bm r}}{(2 \pi)^{3/2}} |  \psi(t) \rangle.
\label{Ffull}
\ee
The expressions in \eq{x-space}, \eq{dipole-KH} and \eq{Ffull} form the starting point for a discussion of the physical picture of the HHG process by infrared laser pulses in the KH frame. These equations are exact within the dipole approximation and can be used together with \eq{signal}, \eq{D} and \eq{dipole} to determine the HHG spectrum for a given system and laser pulse. 

In ab initio approaches based on a grid representation of the spatial variable, the accelerating frame poses a challenge since a spline procedure would have to be implemented for the evaluation of the potential $V(\bm r + \frac{q}{m}\bm \alpha(t))$ at instants of time where the argument $\bm r + \frac{q}{m}\bm \alpha(t)$ is not at a grid point. Moreover a possible divergence of the potential would move in space as a function of time and not be confined to a single point; see the potential in \eq{H}.  Therefore, in this paper, the focus is on an analysis in terms of a typical SFA time-evolution-operator-based perturbative series as presented in the next section.

\subsection{SFA series for the state $| \psi (t) \rangle$}
\label{SFAforPsi}
To develop a perturbative series based on the time-evolution operator formalism, two alternative partitions of the Hamiltonian in \eq{H} are considered. In \eq{H}  one may think of $V (\bm r + \frac{q}{m} \bm \alpha(t) )$ as a perturbation added to the kinetic energy operator of a free electron. The latter kinetic energy part is denoted by $H_V(t) = \bm p^2 /(2m)$, such that one partition is
\be
H(t) = H_V(t) + V(\bm r + \frac{q}{m} \bm \alpha (t) ).
\label{HsplitV}
\ee
The subscript $V$ on the first kinetic energy part in \eq{HsplitV} should remind the reader that in the KH frame, the Volkov states are simply plane wave states solving the TDSE for $H_V(t)$.

The other partition of the Hamiltonian in \eq{H} reads
\be
H(t) = H_0 + V_L(t),
\label{Hsplit1}
\ee
where the potential $V(\bm r)$ has been added and subtracted to form the field-free Hamiltonian
\be
H_0 = \frac{\bm p^2}{2m} + V(\bm r ).
\label{H0}
\ee
The laser-induced interaction in the KH frame reads
\be
V_L(t) =  V(\bm r + \frac{q}{m} \bm \alpha(t)) - V(\bm r).
\label{KH-potential}
\ee
In \eq{Hsplit1},  $V_L(t)$ takes the role as an additional interaction which is added to the field-free Hamiltonian and drives the system away from the field-free initial state, which at the initial time $t_0$ is denoted by $| \psi_0(t_0)\rangle$.

The different partitions in \eq{HsplitV} and \eq{Hsplit1} are used to develop a perturbative series in the potential $V(\bm r)$ for the  time-evolution operator $U(t,t_0)$. Such a perturbative development is the foundation of the SFA; the external field is strong compared with the potential once the electron is freed.  The full time-evolution operator for $H(t)$ satisfies
\be
i \hbar \frac{d}{dt} U(t,t_0) = H(t) U(t,t_0).
\label{teo}
\ee
The time-evolution operator for $H_0$ of \eq{H0}, $U_0(t, t_0)$, and the time-evolution operator for the free-particle Hamiltonian $H_V(t)$ of \eq{HsplitV}, $U_V(t, t_0)$, are defined by equations similar to \eq{teo}. The time-evolution operator of \eq{teo} generates the state $| \psi(t)\rangle$ at a time $t$ from the initial state $ | \psi_0(t_0)\rangle$ at $t_0$ by the equation
\be
|\psi(t) \rangle = U(t,t_0) | \psi_0(t_0) \rangle.
\ee
It is readily shown by application of the Leibniz integral rule that both the operator
\be
U(t,t_0) = U_0(t,t_0) - i \int_{t_0}^t dt' U(t,t') V_L(t') U_0(t',t_0),
\label{Tpart1}
\ee
and the operator
\be
U(t,t_0) = U_V(t,t_0) - i \int_{t_0}^t dt' U(t,t') V U_V(t',t_0),
\label{Tpart2}
\ee
are solutions to \eq{teo}. A Dyson series in the potential $V$ is obtained by iteratively inserting \eq{Tpart2} into \eq{Tpart1}. The present discussion focusses on the leading-order contribution, and to lowest-order in the potential the approximate result for the time-evolution operator reads
\be
U(t,t_0) \simeq U_0(t,t_0) - i \int_{t_0}^t dt' U_V(t,t') V_L(t') U_0(t',t_0),
\label{Tapprox}
\ee
with the associated approximate solution for the state
\be
| \psi(t) \rangle \simeq | \psi_0 (t)\rangle-i \int_{t_0}^t dt' U_V(t,t') V_L(t')  | \psi_0 (t')\rangle. 
\label{psit}
\ee
As expected this equation formally has the same structure as the equation used for the state in the SFA when formulated in the VG or LG - it is just the explicit forms of the operators $U_V(t, t_0)$ and $V_L(t)$ that have changed due to their gauge dependence.

The initial state at time $t$ is given by 
\be
| \psi_0 (t) \rangle =| \psi_0\rangle e^{-\frac{i}{\hbar} E_0 (t-t_0)}, 
\ee
and the $U_V(t, t_0)$ time-evolution operator is explicitly given by 
\be
\label{U_V}
U_V(t,t_0)= e^{-\frac{i}{\hbar} \frac{\bm p^2}{2m} (t-t_0)} = \int d \bm k  | \bm k \rangle \langle \bm k | e^{-\frac{i}{\hbar} 
\frac{\hbar^2 \bm k^2}{2m}(t-t_0)},
\ee
with 
\be
\langle \bm r | \bm k  \rangle  = \frac{1}{(2\pi)^{3/2}} e^{i \bm k \cdot \bm r}.
\ee
In subsequent formulas it will be convenient to introduce the following short-hand notation for the state in \eq{psit}
\be
|\psi (t) \rangle \simeq | \psi_0(t)\rangle + |\psi_1(t) \rangle
\label{psi_sh}
\ee
with $| \psi_1(t) \rangle = -i \int_{t_0}^t dt' U_V(t,t') V_L(t')  | \psi_0 (t') \rangle.$ The state $| \psi_1(t) \rangle$ describes the laser-induced transition from the ground state and into the continuum.

\subsection{SFA for the dipole acceleration in the KH frame}
\label{DipoleaccKH}
In this section, SFA expressions for the dipole acceleration in the KH frame are considered. 

If \eq{x-space} is taken as the starting point, and \eq{psi_sh} is inserted, the result for the dipole acceleration reads
\begin{eqnarray}
\label{expan}
\frac{d^2}{dt^2} \langle  \bm D  \rangle &\simeq& q  \langle \psi_0 (t)| - \nabla_{\bm r}  V(\bm r + \frac{q}{m} \bm \alpha(t)) | \psi_0(t)\rangle, \nonumber \\
&+& (q  \langle \psi_0(t) | - \nabla_{\bm r}  V(\bm r + \frac{q}{m} \bm \alpha(t)) | \psi_1(t)\rangle \nonumber \\ 
&+& c.c.).
\end{eqnarray}
where $c.c.$ denotes complex conjugation of the first term in the parenthesis and where $q \langle \psi_1(t) |  - \nabla_{\bm r}  V(\bm r + \frac{q}{m} \bm \alpha(t)) | \psi_1(t) \rangle$ and higher-order terms have been neglected. The time-dependence associated with the evolution in the initial state cancels in the first, leading-order term in \eq{expan}. This leading-order contribution can be re-written as
\begin{eqnarray}
\label{x0}
\frac{d^2}{dt^2} \langle  \bm D  \rangle_0  &=& q \int d\bm r \rho_0(\bm r)  \left( - \nabla_{\bm r}  V(\bm r + \frac{q}{m} \bm \alpha(t) ) \right) \nonumber \\ 
&=& q \int d\bm r \rho_0(\bm r-\frac{q}{m} \bm \alpha(t))  \left( - \nabla_{\bm r}  V(\bm r) \right),
\end{eqnarray}
where $\rho_0(\bm r) = | \psi_0(\bm r)|^2$ is the initial-state probability density, and where the second line follows from the first by a simple change of coordinate.  In terms of interpretation, the first line in \eq{x0} emphasizes the probing of the static charge density by the force of the potential that changes its origin due to the laser-induced quiver motion. The second line of \eq{x0},  on the other hand, emphasizes the probing of the force of the static potential by the quivering charge density set in motion by the external laser field.

\begin{figure}
\includegraphics[width=0.45\textwidth]{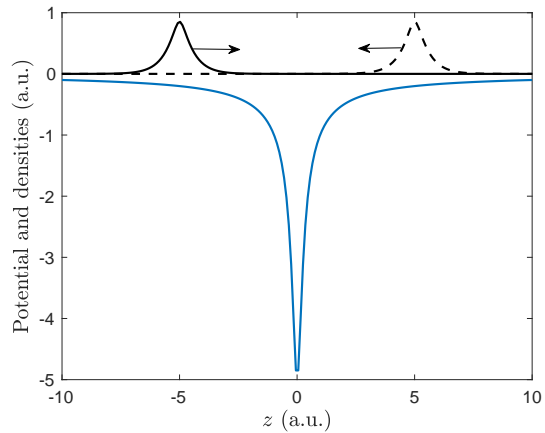}
\caption{Illustration of elements entering the integrant of \eq{x0} at two different instants of time with opposite directions of $\bm \alpha$. The figure shows the Coulomb potential (blue, lower curve), shifted hydrogenic ground state densities (multiplied by a factor of 4 for clarity) for $\alpha=\pm 5$ a.u., and the direction of the force from the Coulomb potential (arrows). The linear laser polarization and therefore also $\bm \alpha$ is along the $z$ axis. A finite value of $x=x_0=0.2$ was used to avoid the singularity of the potential at the origin. Atomic units (a.u.) are used as units on both axes.}
\label{fig1}
\end{figure}

 Figure \ref{fig1} illustrates the potential and the spatially-shifted densities in the case of atomic hydrogen in its ground state. The densities are shown for two different instants of time with $\bm \alpha$ pointing in opposite directions. Therefore the densities shown by full black and dashed black curves are shifted to opposite directions. The two arrows indicate that the direction of the force on the density changes depending on whether the density moves to the right or to the left in the figure. 

For typical atomic  potentials, the last factor in \eq{x0} is relatively strongly peaked at  the position of the nucleus (origin) while the  density is relatively slowly varying. 
These observations lead to the expectation that the integral can be accurately evaluated in the peaking approximation (PA). In the PA, the integral over a product of a strongly peaked function, $g(x)$, and a slowly varying function, $f(x)$, is approximated by the product of the slowly varying function evaluated at the argument $x_0$ for the peak at $g(x_0)$ and the integral over the peaked function, i.e., $\int dx f(x) g(x) \simeq f(x_0) \int dx g(x)$. The PA has found numerous applications in evaluation of difficult integrals in scattering theory and atomic collisions, see, e.g., Ref.~\cite{Kyle1969}.  Due to the simplifications implied by the PA, it is interesting to consider this approach in the present case, where the integrant has some properties in favor for its application. Setting aside mathematic rigor and scaling away issues related to the presence of a factor $\left( \int d \bm r \nabla_{\bm r} V(\bm r) \right)$, the time-dependent part responsible for the  HHG in the PA is modelled by the following expression in the direction of the linear polariztion $\bm \epsilon$
\be
\frac{d^2}{dt^2} \langle  \bm \epsilon \cdot \bm D  \rangle^\text{PA}_0  \sim - q \,  \text{sgn} (\alpha(t)) \, \rho_0(- \frac{q}{m} \bm \alpha(t) ) .
\label{x0peaking}
\ee
The factor $\text{sgn}(\alpha(t))$ accounts for the sign of the quiver amplitude as a function of time. Physically, the alternating sign describes that the force on the electron changes direction depending on whether $\alpha(t)$ is positive or negative as is seen from \eq{x0}: The initial-state density quivers back and forth in the force-field from the stationary potential [see also Fig.~\ref{fig1}].  Equation \eqref{x0peaking} shows that at the level of the PA, harmonics are generated from the time-dependent density $\rho_0(- \frac{q}{m} \bm \alpha(t) )$.  Hence, in the PA, the leading-order term in the SFA for HHG in the KH frame generates harmonics as a consequence of the laser-induced quiver motion traversing  the density in the initial state. Although physically appealing, this PA approach is too simple since the combination of the density and the potential term in \eq{x0} is needed for the correct behavior of the spectra.  These aspects are discussed further below in Sec.~\ref{Example}.  If one goes back to \eq{x0}, this leading-order term leads to a physical picture where the harmonics are generated when the laser-induced space-dependent probing of the density is weighted by the local force. Or, alternatively, when the force from the oscillating potential probes the density in the initial state. Clearly, the laser-induced time-dependent probing of the spatial density of the initial state is offering a different perspective on the mechanism of   HHG than the three-step model of SFA in the LG or VG.

\subsubsection{The time-dependent inelastic form factor $F(\bm k, t)$}
\label{Fstuff}
If \eq{dipole-KH} is taken as the starting point for a discussion of the dipole acceleration in the KH frame, the implications of performing the SFA of \eq{psit} (or \eq{psi_sh} when using the short-hand notation) for the state $| \psi(t) \rangle$ enter at the level of the evaluation of the time-dependent form factor of \eq{Ffull}. The approximation in \eq{psit} gives
\be
F(\bm k, t) \simeq F_0(\bm k) + F_{01}(\bm k, t) + F_{01}(- \bm k, t)^*
\label{Fexpan}
\ee
with the time-independent form factor
\be
F_0(\bm k ) = \langle \psi_0 | \frac{e^{i \bm k \cdot \bm r}}{(2 \pi)^{3/2}} | \psi_0 \rangle, 
\ee
and the time-dependent form factor
\be
F_{01}(\bm k, t ) = \langle \psi_0(t)  |  \frac{e^{i \bm k \cdot \bm r}}{(2 \pi)^{3/2}} | \psi_1(t) \rangle. 
\ee
It is seen that the SFA for HHG in the KH frame gives a contribution which is proportional to the density in the initial state through
\be
F_0(\bm k) =  \frac{1}{(2 \pi)^{3/2}}\int d \bm r \rho_0(\bm r) e^{i \bm k \cdot \bm r},
\label{F0}
\ee
with $\rho_0(\bm r)$ the initial-state density in $\bm r$-space. The presence of a leading-order term proportional to the density is a difference compared to the SFA model for HHG formulated in the LG or VG, where the evaluation of the dipole acceleration amplitude involves a consideration of products of ionization and recombination matrix elements to and from the Volkov continuum~\cite{Lewenstein1994}; see also Sec.~\ref{Remarks}.

If \eq{F0} is inserted in \eq{dipole-KH} the leading-order contribution to the dipole acceleration in the KH frame SFA for HHG is given by
\be
\frac{d^2}{dt^2} \langle  \bm D  \rangle_0 = -i q\int  d \bm k  \bm k \tilde{V}(\bm k)  F_0(\bm k)  e^{i \frac{q}{m} \bm k \cdot \bm \alpha(t) } .
\label{dipole-KH0}
\ee
Of course the physical interpretation of \eq{dipole-KH0} is the same as the interpretation of the first term in  \eq{expan}, i.e., in terms of the expectation value of the force from the oscillating potential on the initial state.  For example, in $\bm k$-space, the first factor linear in the vector $\bm k$ inside the integral in \eq{dipole-KH0} comes from the force term. 

Equation \eqref{dipole-KH0} shows clearly how the selection rules for the harmonics come from a combination of the symmetry properties of the potential $\tilde{V}(\bm k)$ and the elastic form factor $F_0(\bm k)$. If the target is isotropic, the product $ \bm k \tilde V(\bm k)  F_0(\bm k)$ is an odd function of $\bm k$, and the last time-dependent factor in \eq{dipole-KH0} will lead to the usual odd-even selection rules; see the next section. Equation \eqref{dipole-KH0} is attractive for numerical evaluation and will be used in the example of Sec.~\ref{Example}.

\subsection{Long-pulse limit}
\label{LongPulse}
The limit of an infinitely  long pulse allows further analysis of \eq{dipole-KH}. The following form is considered for the time-dependent quiver motion
\be
\bm \alpha(t) = \bm \alpha_0 \sin(\omega_L t),
\label{long-pulse}
\ee
with $\bm \alpha_0$ the time-independent amplitude and $\omega_L$ the frequency of the driving field.  With this choice, the last factor in \eq{dipole-KH0} is re-expressed in terms of Bessel functions of integer order, $J_n(x)$, using the Jacobi-Anger relation
\be
e^{i \frac{q}{m} \bm \alpha_0 \cdot \bm k \sin(\omega_L t)} = \sum_n J_n(\frac{q}{m} \bm \alpha_0 \cdot \bm k) e^{i n \omega_L t}.
\label{expo-single}
\ee
Inserting \eq{expo-single} into \eq{dipole-KH0} gives
\be
\frac{d^2}{dt^2} \langle  \bm D  \rangle = -i q \sum_n e^{i n \omega_L t}  \int  d \bm k J_n(\frac{q}{m}\bm \alpha_0 \cdot \bm k)   \bm k \tilde V(\bm k)  F(\bm k, t) 
\label{long-pulse-signal1}
\ee
To relate to the spectrum, the time-integral in \eq{D} is considered. If the time-dependence in $F(\bm k, t) $ can be neglected either because $F(\bm k, t) \simeq 1$ due to the behaviour of the integrant in \eq{Ffull} or because, the leading-order expression $F_0(\bm k)$ of \eq{F0} is used,  this time integral using the expression in \eq{long-pulse-signal1} gives a delta function, $\int_{-\infty}^\infty dt e^{i x t} = 2\pi \delta(x)$.  In the case of the leading-order approximation, the explicit expression for the HHG amplitude then reads 
\be
\ddot{\tilde{\bm D}}_0(N\omega_L) = -i q \int  d \bm k J_N(\frac{q}{m}\bm \alpha_0 \cdot \bm k)   \bm k \tilde V(\bm k)  F_0(\bm k) 
\label{HHG-amplitude}
\ee
If the target is isotropic,  the integrant is odd when $N$ is even ($J_n(-x) = (-1)^n J_n(x)$) and there are no harmonics generated. Accordingly,  for isotropic targets only odd harmonics can be generated. It is re-assuring that this well-known result follows from the KH frame formulation.

Note in passing, that if the same isotropic system is driven by a two-color field with frequencies $\omega$ and $2 \omega$, $\bm \alpha(t) = \bm \alpha_{01} \sin (\omega_L t) + \bm \alpha_{01} \sin(2 \omega_L t+\phi)$, the exponential in \eq{expo-single} reads
\be
e^{i \frac{q}{m}\bm \alpha(t) \cdot \bm k} = \sum_{n,l} J_n(\frac{q}{m} \bm \alpha_{01} \cdot \bm k) J_l(\frac{q}{m} \bm \alpha_{01} \cdot \bm k) e^{i l \phi} e^{i (n+2l) \omega_L t}.
\ee
The HHG amplitude in this case therefore reads 
 \begin{eqnarray}
 \label{Twocolor}
\ddot{\tilde{\bm D}}_0(N\omega_L) &=& -i q \sum_l \int  d \bm k  J_{N+2l}(\frac{q}{m}\bm \alpha_{01} \cdot \bm k)  \\ \nonumber
&\times & J_l(\frac{q}{m} \bm \alpha_{02} \cdot \bm k)  e^{i l \phi} \bm k \tilde V(\bm k) F_0(\bm k). 
\end{eqnarray}
Equation \eqref{Twocolor} shows that both even and odd harmonics can be generated. When $N$ is odd, $l$ even can give a nonvanishing integrant. When $N$ is even, $l$ odd can give a nonvanishing integrant. 

Finally in this section some relations to the analysis performed in Ref.~\cite{Lakhotia2020} are discussed. In that work HHG spectra of crystalline magnesium fluoride and calcium fluoride obtained with near-infrared fields were  used to reconstruct valence electron densities and effective potentials. The expression  for the HHG amplitude used in Ref.~\cite{Lakhotia2020} for the analysis of the spectra is proportional to the right-hand side of \eq{HHG-amplitude}, except for the absence of the form factor $F_0(\bm k)$ in the formula used in Ref.~\cite{Lakhotia2020}, and with the modification that $\bm k$ takes the role of the crystal momentum; see Eq.~(3) in Ref.~\cite{Lakhotia2020}. The success of the analysis in Ref.~\cite{Lakhotia2020} seems to indicate that the one-dimensional cuts considered in that work for the reconstruction lead to a decrease in the sensitivity to any variation in the form factor with crystal momentum. In any case, the physical perspective implied by \eq{HHG-amplitude} allows an interpretation of spectra that is different than that naturally linked to any three-step model-like picture, be it in atoms, molecules or solids.

\subsection{Remarks on the relation between the leading-order KH frame SFA for HHG and the three-step Lewenstein model}
\label{Remarks}

To shed some light on  the relation between the leading-order KH frame dipole acceleration [\eq{x0} or first term on the right-hand side of \eq{expan}] and the matrix elements of the three-step model, it is useful to re-express the former as
\begin{eqnarray}
\label{KH-VG1}
&&q  \langle \psi_0 (t)| - \nabla_{\bm r}  V(\bm r + \frac{q}{m} \bm \alpha(t)) | \psi_0(t)\rangle
=\\ \nonumber
& &q  (\langle \psi_0 (t) | T(t))| (- \nabla_{\bm r}  V(\bm r)) | (T(t)^\dagger  | \psi_0(t)\rangle),
\end{eqnarray}
with $T(t)$ given by \eq{T}. In \eq{KH-VG1}, the operator $(-\nabla V(\bm r))$ is on the form that is used in the LG or in the VG, i.e., without any instantaneous displacement of the coordinate of the charge particle. Hence, to compare with the expectation value considered in the three-step model emerging from the LG and VG formulations, the state $T(t)^\dagger | \psi_0(t) \rangle$ is analyzed in terms of state-content that enters the SFA for the time-dependent state in the LG or VG. To make this connection, the SFA for the three-step model is briefly summarized. Here, the VG is considered. The LG results are obtained by a simple substitution of the VG expressions by the ones following from a treatment in the LG. In the case of the VG, the theory of Sec.~\ref{SFAforPsi}.  gives the SFA approximation for the state as
\be
| \psi^\text{VG}(t) \rangle \simeq |\psi_0(t) \rangle +  |\psi_1^\text{VG}(t) \rangle.
\label{psi-VG}
\ee
with $|\psi_0(t) \rangle$ the initial state as before and with 
\begin{eqnarray}
 \label{psi1-VG}
 |\psi_1^\text{VG}(t) \rangle= -i \int_{t_0}^t && dt' \int d{\bm k}
 | \psi^{V, \bm k}_\text{VG} (t) \rangle  \\ \nonumber
 &\times& \langle \psi^{V, \bm k}_\text{VG} (t') | V_L^\text{VG}(t') | \psi_0(t') \rangle.
\end{eqnarray}
Here $V_L^\text{VG}(t)= (\bm p - q \bm A(t))^2/(2 m) - \bm p^2/(2 m)$ is the laser-matter operator in the  VG. The time-evolution operator describing the evolution in the laser-dressed continuum is expressed in terms of the VG Volkov states $ \langle \bm r |  \psi^{V, \bm k}_\text{VG} (t) \rangle = (2 \pi)^{-(3/2)} e^{i \bm k \cdot \bm r} e^{-\frac{i}{\hbar} \int_{t_0}^t dt' (\hbar \bm k - q \bm A(t))^2/(2m)}$, which solve the TDSE for a free electron in the presence of $V_L^\text{VG}(t)$. The dominating contributions to the dipole acceleration in the VG are therefore 
\begin{eqnarray}
\label{dipole-VG}
&&q \langle \psi^\text{VG}(t) | - \nabla_{\bm r} V(\bm r) | \psi^\text{VG}(t) \rangle \simeq \\ \nonumber
&& q [\langle \psi_0(t) | - \nabla_{\bm r} V(\bm r) | \psi_1^\text{VG}(t) \rangle  + c.c ],
\end{eqnarray}
where $c.c.$ denotes the complex conjugate of the first term in the square braket. The characteristic three-step physics in terms of strong-field ionization, propagation and recombination steps, is  readily identified from  \eq{dipole-VG} when \eq{psi1-VG} is used.

With these equations at hand the relation between the leading-order KH frame SFA and the SFA of the three-step model can be explored. The state $T(t)^\dagger | \psi_0(t)\rangle $can be formally expressed in terms of its $| \psi_0(t) \rangle$, $ | \psi_1^\text{VG}(t) \rangle$ content and a rest, $| \psi'(t) \rangle$ as
\be
T(t)^\dagger | \psi_0(t) \rangle = c_0(t) | \psi_0(t) \rangle + c_1(t) | \psi_1^\text{VG}(t) \rangle + | \psi'(t) \rangle
\label{KHinVG}
\ee
with 
\be
c_0(t) = \langle \psi_0(t) |T(t)^\dagger  |\psi_0(t) \rangle = \int d \bm r \psi_0(\bm r)^* \psi_0(\bm r - \frac{q}{m} \bm \alpha(t) )
\label{c0}
\ee
and 
\be
c_1(t) = \langle \psi_1^\text{VG}(t) | T(t)^\dagger | \psi_0(t) \rangle.
\label{c1}
\ee
The explicit expression for $c_1(t)$ is not needed here. It can be obtained by using \eq{psi1-VG}. The important point is that $c_1(t) \ne 0$. The expression for $c_0(t)$ clearly shows that this amplitude decreases with increasing $|\alpha(t)|$. 

Inserting \eq{KHinVG} into \eq{KH-VG1} gives for nonpolar systems, where $\langle \psi_0(t) | -\nabla V(\bm r) | \psi_0(t) \rangle =0$ due to parity selection rules,
\begin{eqnarray}
\label{KH-VG2}
& &q  (\langle \psi_0 (t) | T(t))| (- \nabla_{\bm r}  V(\bm r)) | (T(t)^\dagger | \psi_0(t)\rangle) \simeq \\ \nonumber 
&& q [ c_0(t)^* c_1(t) \langle \psi_0(t) | -\nabla V(\bm r) | \psi_1^\text{VG}(t) \rangle \\ \nonumber 
&& + c_0(t) c_1^*(t) \langle \psi_1^\text{VG}(t) | -\nabla V(\bm r) | \psi_0(t) \rangle   \\ \nonumber
&& + c_0(t)^* \langle \psi_0(t) | -\nabla V(\bm r) | \psi'(t) \rangle \\ \nonumber
&& + c_0(t) \langle \psi'(t) | -\nabla V(\bm r) | \psi_0(t)) \rangle \\ \nonumber
&& + c_1(t)^* \langle \psi_1^\text{VG}(t) | - \nabla V(\bm r) | \psi'(t) \rangle \\ \nonumber
&& + c_1(t) \langle \psi'(t) | - \nabla V(\bm r) | \psi_1^\text{VG}(t) \rangle \\ \nonumber
&& + \langle \psi'(t) | - \nabla V(\bm r) | \psi'(t) \rangle \\ \nonumber
&& + |c_1(t)|^2 \langle \psi_1^\text{VG}(t) | - \nabla V(\bm r) | \psi_1^\text{VG}(t) \rangle ].
\end{eqnarray}
Equation \eqref{KH-VG2} shows through the presence of the two first terms on the right-hand side that the leading-order SFA for HHG in the KH frame includes contributions from the matrix elements that enter the three-step model of HHG, see \eq{dipole-VG}.  While \eq{KH-VG2} is solely used for analysis of the nature of the leading-order SFA HHG amplitude in the KH frame and not for actual computation, it is still interesting to note that the amplitude $c_1(t)$ will be of the order of the three-step SFA HHG amplitude. The amplitude $c_0(t)$ can be much larger depending on the value of $\bm \alpha(t)$. It is therefore expected that the largest amplitudes in \eq{KH-VG2} are $c_0(t)^* \langle \psi_0(t) | -\nabla V(\bm r) | \psi'(t) \rangle$ and its complex conjugate, where $|\psi'(t) \rangle$ represent the content of $\langle \bm r | T(t)^\dagger |\psi_0(t) \rangle = \psi_0(\bm r - \frac{q}{m} \bm \alpha(t) , t) $ that is neither in $|\psi_0(t) \rangle$ nor $| \psi_1^\text{VG}(t) \rangle$, i.e., content which in terms of field-free basis states includes excited and continuum states for the potential  $V$.

Finally, in this section, which addresses certain characteristics of the SFA formulation in the KH frame,  the question of the usefulness of a formulation using adiabatic KH frame states is considered. The adiabatic states in the KH frame are discussed in detail in Ref.~\cite{Madsen2021a}. For example, the adiabatic state of the initial state is given by $\psi_0(\bm r + \frac{q}{m} \bm \alpha(t))$, i.e., by the initial state that adiabatically follows and adjusts to the instantaneous value of the position of the charged particle. If this state is used in \eq{KH-VG1} it is seen that the result reduces to the time-independent quantity $q \int d \bm r | \psi_0(\bm r) | ^2 (- \nabla V(\bm r)) $. Hence no harmonics would be generated by that term in the adiabatic approach. Then, not surprisingly, in an approach based on expansion in adiabatic states, the harmonics are generated by nonadiabatic transitions. The nonadiabatic transitions in the KH frame are given by VG matrix elements~\cite{Madsen2021a}. Therefore no benefits, compared to an approach based on a formulation directly in the VG, would emerge in the analysis of HHG from working in the basis of adiabatic KH states.

\subsection{Example}
\label{Example}
In this section, the theory is illustrated by a simple example. The driving laser pulse is taken to be linearly polarized and to be described by  the form
\be
\bm \alpha(t) = \bm \alpha_0 \sin^2(\pi t/T) \sin(\omega_L t),
\label{pulse}
\ee
with $T$ the pulse duration and $\omega_L$ the angular frequency.  The peak amplitude  $\bm \alpha_0 = \alpha_0 \hat{\bm z}$ corresponds to an intensity of  $3.16 \times 10^{13}$ W/cm$^2$. The wavelengths considered will be 800 nm, 1600 nm and 3200 nm as specified in the captions of the figures. For the simplest target of atomic hydrogen, the Fourier transform of the  Coulomb potential, $V(\bm r) = - Z q^2/(4 \pi \epsilon_0 r)$, is readily obtained as 
\be
\tilde V(\bm k ) = -\frac{Zq^2}{4\pi \epsilon_0} \frac{4\pi}{k^2} \frac{1}{(2 \pi)^{3/2}}. 
\label{Vk-hydrogen}
\ee
Using the analytical hydrogenic ground state wave function as the initial state, the form factor is readily evaluated 
\be
F_0(\bm k)  =
\frac{16}{(4 + (k a_0/Z)^2)^2} \frac{1}{(2 \pi)^{3/2}},
\label{F0-hydrogen}
\ee
with $a_0$ the Bohr radius. For hydrogen $Z=1$. The dependence on $Z$ is kept to expose the sensitivity to the nuclear charge. Inserting \eq{Vk-hydrogen} and \eq{F0-hydrogen} into \eq{dipole-KH} allows the evaluation of the dipole acceleration. The spectrum in the polarization direction $\bm \epsilon = \hat{\bm z}$ is obtained from  \eq{D} and \eq{signal}. Alternatively, \eq{Vk-hydrogen} and \eq{F0-hydrogen} can be used in \eq{HHG-amplitude} to obtain the HHG amplitude in the long-pulse limit. 

\begin{figure}
\includegraphics[width=0.45\textwidth]{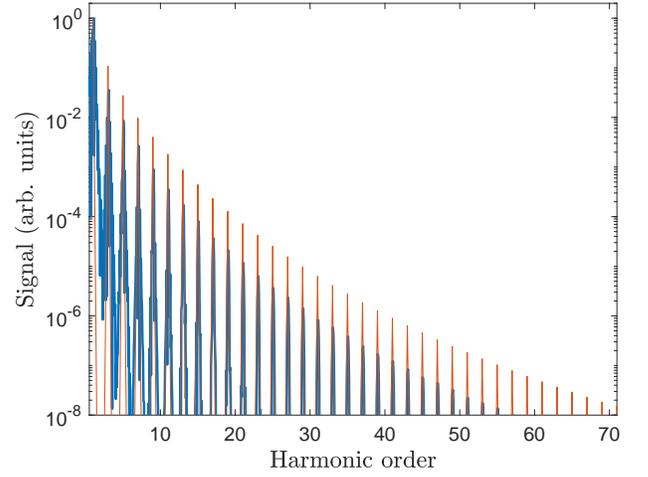}
\caption{HHG spectrum for atomic hydrogen subject to a linearly polarized laser pulse containing 10 cycles at 800 nm and having a peak intensity of  $3.16 \times 10^{13}$ W/cm$^2$. The thicker, blue curve is the result for the leading-order SFA in the  KH frame formulation [\eq{x0} or \eq{dipole-KH0}] for a pulse described by \eq{pulse}. The other thinner, red,  higher-lying curve presents the estimate for the HHG yield in the infinitely long pulse limit based on 
\eq{HHG-amplitude}. The spectra have been nomalized by their maximum values.}
\label{fig2}
\end{figure}

It is noteworthy that the dipole acceleration of \eq{x0} can be evaluated in analytically closed form for the considered hydrogenic example. The result in the linear polarization direction, $\hat{\bm z}$, of the driving pulse reads 
\be
\frac{d^2}{dt^2} \langle  D_z  \rangle_0  = q \, \text{sgn}(\alpha(t)) \, \frac{m^2}{\alpha(t)^2 q^2}  
\{ 1 + 
e^{-2 \tilde \alpha  }   [- 2 \tilde \alpha  ( \tilde \alpha   + 1) -1] \},
\label{x0-exact}
\ee
where a shorthand notation with $\tilde \alpha  = | \frac{q}{m} \alpha (t) | \frac{Z}{a_0}$ was used. The result using the PA of \eq{x0peaking} reads
\be
\frac{d^2}{dt^2} \langle  D_z  \rangle^\text{PA}_0 \sim
-q \,2 \, \text{sgn}(\alpha(t) ) \,e^{-2 \tilde \alpha}.
\label{PA}
\ee
As in Sec.~\ref{DipoleaccKH}, this result shows that the source of the dipole acceleration of HHG can be interpreted as  stemming from laser-induced oscillations of the initial state density $\rho_0(\tilde \alpha) = \frac{1}{\pi} e^{- 2 \tilde \alpha}$ in the force field from the atomic potential.
\begin{figure}
\includegraphics[width=0.45\textwidth]{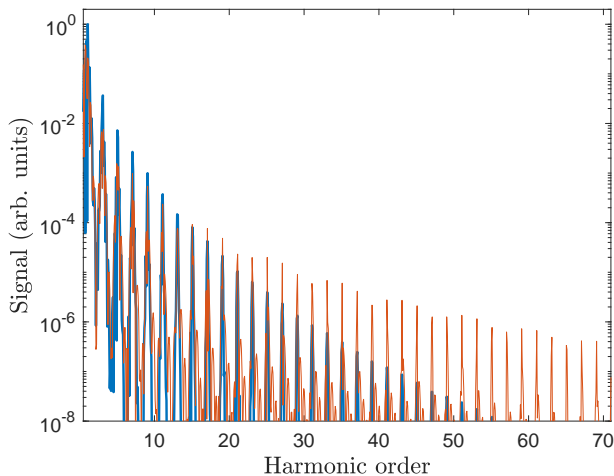}
\caption{ HHG spectrum for atomic hydrogen subject to a linearly polarized laser pulse containing 10 cycles at 800 nm and having a peak intensity of  $3.16 \times 10^{13}$ W/cm$^2$. The thicker, blue curve is the result for the leading-order SFA in the  KH frame formulation as in Fig.~\ref{fig2}. The other thinner, red curve is the result of the PA of \eq{PA} [see also \eq{x0peaking}].}
\label{fig3}
\end{figure}

Figure~\ref{fig2} shows results for the HHG spectra in atomic hydrogen for a pulse with 10 cycles (lower-lying, blue curve) as well as results for the  infinitely long pulse (higher-lying, red curve). It is seen from the figure that the leading-order term in the KH frame SFA for HHG gives the characteristic odd harmonics.  The plateau and cut-off features that are often seen in modelling with the three-step model, but often less clearly observed in experimental data, are not captured by the leading-order KH term. Since, however, the KH leading-order contribution is related to the dominating $|\psi_0(t) \rangle$ part of the state in \eq{psit} or \eq{psi_sh}, one could expect that the lower-order harmonics (including the sub-threshold harmonics) are relatively accurately described by the KH frame SFA.   In the conventional SFA for HHG, the three-step model is involved in the generation of both above- and sub-threshold harmonics. There is, therefore, no contribution solely depending on the $| \psi_0 (t) \rangle$ part of the propagated state. The leading-order contribution is from terms involving $|\psi_0(t) \rangle$ and the rescattered part, see \eq{dipole-VG}. One could therefore expect the LG and VG formulations of the SFA of HHG, i.e., the quantum version of the three-step model, to be more challenged than the present KH frame formulation in accurately describing the lower-order harmonics; see, e.g., Ref.~\cite{Perez-Hernandez2009}. Since the present  focus is  on the KH frame SFA formulation and the simplest illustration of the theory, the full assessment of these expectations is work for the future. 

The ad hoc PA of  \eq{x0peaking} is very easily evaluated. Only knowledge of the laser pulse and the  density of the initial state orbital is needed for an estimation of the HHG spectrum in this very naive modelling. The density can be obtained for many systems using standard quantum chemistry or solid-state software packages. Due to its attractiveness for applications, it is therefore, despite its crudeness,  relevant to assess  the accuracy of the PA of \eq{x0peaking}. In the case of hydrogen initially in its ground state, the result for the PA is given in \eq{PA}. In Fig.~\ref{fig3}, the spectrum of the leading-order SFA in the KH frame, also shown in Fig.~\ref{fig2}, is compared with the result for the HHG following the use of the PA for the dipole acceleration. The figure shows that the PA captures the presence of only odd harmonics. The figure shows quantitative  disagreement between the spectrum generated by the leading-order term of the dipole acceleration in the KH frame and its PA result at harmonics of both low and higher order. 
\begin{figure}
\includegraphics[width=0.50\textwidth]{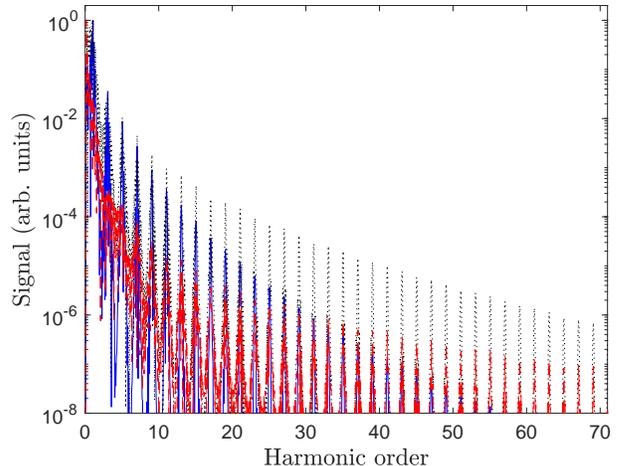}
\caption{Wavelength dependence of the HHG spectrum for atomic hydrogen subject to a linearly polarized laser pulse containing 10 cycles at 800 nm (blue), 1600 nm (black, dotted) and 3200 nm (red, dashed) having a peak intensity of  $3.16 \times 10^{13}$ W/cm$^2$. The curves show the results for the leading-order SFA in the  KH frame formulation  [\eq{x0} or \eq{dipole-KH0}] for a pulse described by \eq{pulse}.}
\label{fig4}
\end{figure}

Figure~\ref{fig4} illustrates the wavelength dependence of the HHG spectra obtained from the leading-order KH frame term. The figure shows that as the wavelength increases the intensity of the higher-order harmonics increase relative to that of the lower orders. For a fixed intensity, the $|\psi_1(t)\rangle$ correction to $| \psi_0(t) \rangle$ in the state in \eq{psi_sh} could be expected to decrease with increasing wavelength since it takes more photons to ionize. Hence, the prediction of the leading-order SFA for HHG in the KH frame could become more accurate with increasing wavelength, again the precise assessment of these expectations is for the future.

\begin{figure}
\includegraphics[width=0.49\textwidth]{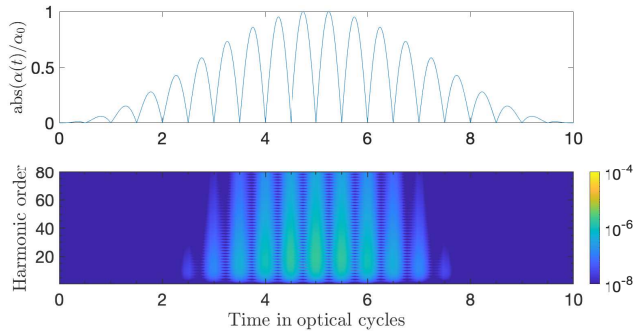}
\caption{Time-frequency analysis of the HHG spectrum from Fig.~\ref{fig2}.  The upper panel shows the absolute value of the normalized quiver amplitude,  \eq{pulse}. The lower panel shows the time-frequency profile in arbitrary units.}
\label{fig5}
\end{figure}

Finally, it is useful to consider a time-frequency analysis of the HHG signal produced by the leading-order term in the KH frame SFA formulation, i.e., from the term in \eq{x0} or in \eq{dipole-KH0}. To this end, a Gabor transform  of the considered term is performed 
\begin{eqnarray}
\label{G}
G_\tau (\omega,t) &=& \int dt'  e^{-i \omega t' -(t-t')^2/(2 \tau^2) } \\ \nonumber
&\times&q \langle \psi_0(t)' | - \nabla_{\bm r} V(\bm r + \frac{q}{m} \bm \alpha (t') | \psi_0(t') \rangle,
\end{eqnarray}
where the width of the time window is chosen to $5 \pi$ a.u. The absolute  value of the quiver amplitude for the pulse used in Fig.~\ref{fig2} is shown in the upper panel of Fig.~\ref{fig5}.  The lower panel of Fig.~\ref{fig5}, shows $| G_\tau(\omega,t)|^2$ from \eq{G}. A comparison of the upper and lower panels shows that most harmonic signal is emitted when the quiver amplitude goes through zero. The dominance of these emission times reflects that the dipole acceleration changes sign at these instants and quickly assumes its maximal values; see the right-hand side of \eq{x0}. In the frame, where the potential is time-dependent and  its minimum moves back and forth according to $\bm r + \frac{q}{m}\bm \alpha(t)$ the maximum of the initial state density is probed for $\bm \alpha(t) =0$ and the direction of the force changes sign, therefore most harmonic signal is emitted at these instants. 

In closing this section on illustrative examples, it is mentioned that  laser pulses of general ellipticity have also been considered. Results that are not illustrated in figures here show  that when the ellipticity increases from linear to circular polarization, the signal in the higher orders in the HHG spectrum decreases. This behavior is well-known from the three-step model.  The description of the physical reason behind this decrease is linked to the model. In the three-step model, the signal decreases because the amplitude for recombination drops since the trajectory of the tunnel-ionized electron misses the parent ion when the ellipticity increases. In the leading-order KH-frame formulation, the signal decreases because the force from the potential [first line in \eq{x0}]  or, from another point of view, the density of the initial state [last line in \eq{x0}] varies more slowly when $\bm \alpha(t)$ approaches a circle.

\section{Conclusion and outlook}
\label{Conclusion}
In this work, the SFA for HHG in infrared and near-infrared laser pulses was investigated based on a formulation in the accelerating KH frame. While the development of the SFA series for the quantum state in the LG or VG leads to a picture where the harmonics are generated by a dominating term describing three-step ionization, propagation and recombination processes, and consequently explicitly involves the consideration of the electronic continuum, the situation in the KH frame is different. Here the leading-order contribution to the HHG process describes the harmonics generated as the force of the time-dependent oscillating potential, $-\nabla V(\bm r + \frac{q}{m} \bm \alpha(t) )$, traverses the initial state field-free density, or equivalently as the time-dependent oscillating density $\rho(\bm r - \frac{q}{m} \bm \alpha(t))$ probes the force field of the time-independent static potential.  This alternative and supplementary perspective has recently proved its usefulness when expressed in $\bm k$-space, as also considered here,  in analysis of HHG from solid-state systems where it facilitated the extraction of information about valence potentials and real-space densities based on harmonics up to order $\simeq 13$~\cite{Lakhotia2020}. The successful retrieval of potentials and densities suggest that the leading-order KH dipole acceleration accurately describe the HHG process in the regime considered in Ref.~\cite{Lakhotia2020}.  In addition to the leading-order term, in the present work, the next term of the state vector was also explicitly given and the procedure for iteration  in the potential of the system was discussed. Some remarks were given on the elements of the three-step model that contribute to the leading-order KH frame dipole acceleration. The theory was illustrated by a simple example in atomic hydrogen. 

In the future, it could be relevant to re-visit predictions for HHG in molecules with the KH frame approach and, e.g., consider spectra as a function of internuclear distance and alignment and orientation with respect to polarization of the  laser pulse. Moreover, regarding molecules, it seems that the physical picture associated with the present formulation in terms of probing of the time-dependent laser-shifted density by the force of the static molecular potential (or equivalently the probing of the laser-displaced force of the molecular potential by the static density) could be useful in investigation of chiral properties due to the very direct link to the chiral potential and the orbitals in the KH frame. By changing $\bm \alpha_0$ for light with nonlinear polarization different parts of the potential and the density will be traced out [see \eq{x0}]. Likewise it seems natural to pursue a formulation and evaluation of the KH frame SFA for strong-field ionization with infrared fields.

\begin{acknowledgments}
Discussions with Simon Vendelbo Bylling Jensen are acknowledged. This work was supported by the Danish Council for Independent Research (Grant No. 9040-00001B).
\end{acknowledgments}


\begin{thebibliography}{16}%
\makeatletter
\providecommand \@ifxundefined [1]{%
 \@ifx{#1\undefined}
}%
\providecommand \@ifnum [1]{%
 \ifnum #1\expandafter \@firstoftwo
 \else \expandafter \@secondoftwo
 \fi
}%
\providecommand \@ifx [1]{%
 \ifx #1\expandafter \@firstoftwo
 \else \expandafter \@secondoftwo
 \fi
}%
\providecommand \natexlab [1]{#1}%
\providecommand \enquote  [1]{``#1''}%
\providecommand \bibnamefont  [1]{#1}%
\providecommand \bibfnamefont [1]{#1}%
\providecommand \citenamefont [1]{#1}%
\providecommand \href@noop [0]{\@secondoftwo}%
\providecommand \href [0]{\begingroup \@sanitize@url \@href}%
\providecommand \@href[1]{\@@startlink{#1}\@@href}%
\providecommand \@@href[1]{\endgroup#1\@@endlink}%
\providecommand \@sanitize@url [0]{\catcode `\\12\catcode `\$12\catcode
  `\&12\catcode `\#12\catcode `\^12\catcode `\_12\catcode `\%12\relax}%
\providecommand \@@startlink[1]{}%
\providecommand \@@endlink[0]{}%
\providecommand \url  [0]{\begingroup\@sanitize@url \@url }%
\providecommand \@url [1]{\endgroup\@href {#1}{\urlprefix }}%
\providecommand \urlprefix  [0]{URL }%
\providecommand \Eprint [0]{\href }%
\providecommand \doibase [0]{https://doi.org/}%
\providecommand \selectlanguage [0]{\@gobble}%
\providecommand \bibinfo  [0]{\@secondoftwo}%
\providecommand \bibfield  [0]{\@secondoftwo}%
\providecommand \translation [1]{[#1]}%
\providecommand \BibitemOpen [0]{}%
\providecommand \bibitemStop [0]{}%
\providecommand \bibitemNoStop [0]{.\EOS\space}%
\providecommand \EOS [0]{\spacefactor3000\relax}%
\providecommand \BibitemShut  [1]{\csname bibitem#1\endcsname}%
\let\auto@bib@innerbib\@empty
\bibitem [{\citenamefont {Schafer}\ \emph {et~al.}(1993)\citenamefont
  {Schafer}, \citenamefont {Yang}, \citenamefont {DiMauro},\ and\ \citenamefont
  {Kulander}}]{Schafer1993}%
  \BibitemOpen
  \bibfield  {author} {\bibinfo {author} {\bibfnamefont {K.~J.}\ \bibnamefont
  {Schafer}}, \bibinfo {author} {\bibfnamefont {B.}~\bibnamefont {Yang}},
  \bibinfo {author} {\bibfnamefont {L.~F.}\ \bibnamefont {DiMauro}},\ and\
  \bibinfo {author} {\bibfnamefont {K.~C.}\ \bibnamefont {Kulander}},\
  }\bibfield  {title} {\bibinfo {title} {Above threshold ionization beyond the
  high harmonic cutoff},\ }\href {https://doi.org/10.1103/PhysRevLett.70.1599}
  {\bibfield  {journal} {\bibinfo  {journal} {Phys. Rev. Lett.}\ }\textbf
  {\bibinfo {volume} {70}},\ \bibinfo {pages} {1599} (\bibinfo {year}
  {1993})}\BibitemShut {NoStop}%
\bibitem [{\citenamefont {Corkum}(1993)}]{Corkum1993}%
  \BibitemOpen
  \bibfield  {author} {\bibinfo {author} {\bibfnamefont {P.~B.}\ \bibnamefont
  {Corkum}},\ }\bibfield  {title} {\bibinfo {title} {Plasma perspective on
  strong field multiphoton ionization},\ }\href
  {https://doi.org/10.1103/PhysRevLett.71.1994} {\bibfield  {journal} {\bibinfo
   {journal} {Phys. Rev. Lett.}\ }\textbf {\bibinfo {volume} {71}},\ \bibinfo
  {pages} {1994} (\bibinfo {year} {1993})}\BibitemShut {NoStop}%
\bibitem [{\citenamefont {Kulander}\ \emph {et~al.}(1993)\citenamefont
  {Kulander}, \citenamefont {Schafer},\ and\ \citenamefont
  {Krause}}]{Kulander1993}%
  \BibitemOpen
  \bibfield  {author} {\bibinfo {author} {\bibfnamefont {K.~C.}\ \bibnamefont
  {Kulander}}, \bibinfo {author} {\bibfnamefont {K.~J.}\ \bibnamefont
  {Schafer}},\ and\ \bibinfo {author} {\bibfnamefont {J.~L.}\ \bibnamefont
  {Krause}},\ }in\ \href@noop {} {\emph {\bibinfo {booktitle} {Super-Intense
  Laser-Atom Physics}}},\ \bibinfo {editor} {edited by\ \bibinfo {editor}
  {\bibfnamefont {B.}~\bibnamefont {Piraux}}, \bibinfo {editor} {\bibfnamefont
  {A.}~\bibnamefont {L’Hullier}},\ and\ \bibinfo {editor} {\bibfnamefont
  {K.}~\bibnamefont {Rzazewski}}}\ (\bibinfo  {publisher} {Plenum, New York},\
  \bibinfo {year} {1993})\BibitemShut {NoStop}%
\bibitem [{\citenamefont {Lewenstein}\ \emph {et~al.}(1994)\citenamefont
  {Lewenstein}, \citenamefont {Balcou}, \citenamefont {Ivanov}, \citenamefont
  {L'Huillier},\ and\ \citenamefont {Corkum}}]{Lewenstein1994}%
  \BibitemOpen
  \bibfield  {author} {\bibinfo {author} {\bibfnamefont {M.}~\bibnamefont
  {Lewenstein}}, \bibinfo {author} {\bibfnamefont {P.}~\bibnamefont {Balcou}},
  \bibinfo {author} {\bibfnamefont {M.~Y.}\ \bibnamefont {Ivanov}}, \bibinfo
  {author} {\bibfnamefont {A.}~\bibnamefont {L'Huillier}},\ and\ \bibinfo
  {author} {\bibfnamefont {P.~B.}\ \bibnamefont {Corkum}},\ }\bibfield  {title}
  {\bibinfo {title} {Theory of high-harmonic generation by low-frequency laser
  fields},\ }\href {https://doi.org/10.1103/PhysRevA.49.2117} {\bibfield
  {journal} {\bibinfo  {journal} {Phys. Rev. A}\ }\textbf {\bibinfo {volume}
  {49}},\ \bibinfo {pages} {2117} (\bibinfo {year} {1994})}\BibitemShut
  {NoStop}%
\bibitem [{\citenamefont {Vampa}\ \emph {et~al.}(2014)\citenamefont {Vampa},
  \citenamefont {McDonald}, \citenamefont {Orlando}, \citenamefont {Klug},
  \citenamefont {Corkum},\ and\ \citenamefont {Brabec}}]{Vampa2014}%
  \BibitemOpen
  \bibfield  {author} {\bibinfo {author} {\bibfnamefont {G.}~\bibnamefont
  {Vampa}}, \bibinfo {author} {\bibfnamefont {C.~R.}\ \bibnamefont {McDonald}},
  \bibinfo {author} {\bibfnamefont {G.}~\bibnamefont {Orlando}}, \bibinfo
  {author} {\bibfnamefont {D.~D.}\ \bibnamefont {Klug}}, \bibinfo {author}
  {\bibfnamefont {P.~B.}\ \bibnamefont {Corkum}},\ and\ \bibinfo {author}
  {\bibfnamefont {T.}~\bibnamefont {Brabec}},\ }\bibfield  {title} {\bibinfo
  {title} {Theoretical analysis of high-harmonic generation in solids},\ }\href
  {https://doi.org/10.1103/PhysRevLett.113.073901} {\bibfield  {journal}
  {\bibinfo  {journal} {Phys. Rev. Lett.}\ }\textbf {\bibinfo {volume} {113}},\
  \bibinfo {pages} {073901} (\bibinfo {year} {2014})}\BibitemShut {NoStop}%
\bibitem [{\citenamefont {Itatani}\ \emph {et~al.}(2004)\citenamefont
  {Itatani}, \citenamefont {Levesque}, \citenamefont {Zeidler}, \citenamefont
  {Niikura}, \citenamefont {P{\'e}pin}, \citenamefont {Kieffer}, \citenamefont
  {Corkum},\ and\ \citenamefont {Villeneuve}}]{Itatani2004}%
  \BibitemOpen
  \bibfield  {author} {\bibinfo {author} {\bibfnamefont {J.}~\bibnamefont
  {Itatani}}, \bibinfo {author} {\bibfnamefont {J.}~\bibnamefont {Levesque}},
  \bibinfo {author} {\bibfnamefont {D.}~\bibnamefont {Zeidler}}, \bibinfo
  {author} {\bibfnamefont {H.}~\bibnamefont {Niikura}}, \bibinfo {author}
  {\bibfnamefont {H.}~\bibnamefont {P{\'e}pin}}, \bibinfo {author}
  {\bibfnamefont {J.~C.}\ \bibnamefont {Kieffer}}, \bibinfo {author}
  {\bibfnamefont {P.~B.}\ \bibnamefont {Corkum}},\ and\ \bibinfo {author}
  {\bibfnamefont {D.~M.}\ \bibnamefont {Villeneuve}},\ }\bibfield  {title}
  {\bibinfo {title} {Tomographic imaging of molecular orbitals},\ }\href
  {https://doi.org/10.1038/nature03183} {\bibfield  {journal} {\bibinfo
  {journal} {Nature}\ }\textbf {\bibinfo {volume} {432}},\ \bibinfo {pages}
  {867} (\bibinfo {year} {2004})}\BibitemShut {NoStop}%
\bibitem [{\citenamefont {Kraus}\ \emph {et~al.}(2015)\citenamefont {Kraus},
  \citenamefont {Mignolet}, \citenamefont {Baykusheva}, \citenamefont
  {Rupenyan}, \citenamefont {Horn{\'y}}, \citenamefont {Penka}, \citenamefont
  {Grassi}, \citenamefont {Tolstikhin}, \citenamefont {Schneider},
  \citenamefont {Jensen}, \citenamefont {Madsen}, \citenamefont {Bandrauk},
  \citenamefont {Remacle},\ and\ \citenamefont {W{\"o}rner}}]{Kraus2015}%
  \BibitemOpen
  \bibfield  {author} {\bibinfo {author} {\bibfnamefont {P.~M.}\ \bibnamefont
  {Kraus}}, \bibinfo {author} {\bibfnamefont {B.}~\bibnamefont {Mignolet}},
  \bibinfo {author} {\bibfnamefont {D.}~\bibnamefont {Baykusheva}}, \bibinfo
  {author} {\bibfnamefont {A.}~\bibnamefont {Rupenyan}}, \bibinfo {author}
  {\bibfnamefont {L.}~\bibnamefont {Horn{\'y}}}, \bibinfo {author}
  {\bibfnamefont {E.~F.}\ \bibnamefont {Penka}}, \bibinfo {author}
  {\bibfnamefont {G.}~\bibnamefont {Grassi}}, \bibinfo {author} {\bibfnamefont
  {O.~I.}\ \bibnamefont {Tolstikhin}}, \bibinfo {author} {\bibfnamefont
  {J.}~\bibnamefont {Schneider}}, \bibinfo {author} {\bibfnamefont
  {F.}~\bibnamefont {Jensen}}, \bibinfo {author} {\bibfnamefont {L.~B.}\
  \bibnamefont {Madsen}}, \bibinfo {author} {\bibfnamefont {A.~D.}\
  \bibnamefont {Bandrauk}}, \bibinfo {author} {\bibfnamefont {F.}~\bibnamefont
  {Remacle}},\ and\ \bibinfo {author} {\bibfnamefont {H.~J.}\ \bibnamefont
  {W{\"o}rner}},\ }\bibfield  {title} {\bibinfo {title} {Measurement and laser
  control of attosecond charge migration in ionized iodoacetylene},\ }\href
  {https://doi.org/10.1126/science.aab2160} {\bibfield  {journal} {\bibinfo
  {journal} {Science}\ }\textbf {\bibinfo {volume} {350}},\ \bibinfo {pages}
  {790} (\bibinfo {year} {2015})}\BibitemShut {NoStop}%
\bibitem [{\citenamefont {Vampa}\ \emph {et~al.}(2015)\citenamefont {Vampa},
  \citenamefont {Hammond}, \citenamefont {Thir\'e}, \citenamefont {Schmidt},
  \citenamefont {L\'egar\'e}, \citenamefont {McDonald}, \citenamefont {Brabec},
  \citenamefont {Klug},\ and\ \citenamefont {Corkum}}]{Vampa2015}%
  \BibitemOpen
  \bibfield  {author} {\bibinfo {author} {\bibfnamefont {G.}~\bibnamefont
  {Vampa}}, \bibinfo {author} {\bibfnamefont {T.~J.}\ \bibnamefont {Hammond}},
  \bibinfo {author} {\bibfnamefont {N.}~\bibnamefont {Thir\'e}}, \bibinfo
  {author} {\bibfnamefont {B.~E.}\ \bibnamefont {Schmidt}}, \bibinfo {author}
  {\bibfnamefont {F.}~\bibnamefont {L\'egar\'e}}, \bibinfo {author}
  {\bibfnamefont {C.~R.}\ \bibnamefont {McDonald}}, \bibinfo {author}
  {\bibfnamefont {T.}~\bibnamefont {Brabec}}, \bibinfo {author} {\bibfnamefont
  {D.~D.}\ \bibnamefont {Klug}},\ and\ \bibinfo {author} {\bibfnamefont
  {P.~B.}\ \bibnamefont {Corkum}},\ }\bibfield  {title} {\bibinfo {title}
  {All-optical reconstruction of crystal band structure},\ }\href
  {https://doi.org/10.1103/PhysRevLett.115.193603} {\bibfield  {journal}
  {\bibinfo  {journal} {Phys. Rev. Lett.}\ }\textbf {\bibinfo {volume} {115}},\
  \bibinfo {pages} {193603} (\bibinfo {year} {2015})}\BibitemShut {NoStop}%
\bibitem [{\citenamefont {P\'{e}rez-Hern\'{a}ndez}\ \emph
  {et~al.}(2009)\citenamefont {P\'{e}rez-Hern\'{a}ndez}, \citenamefont {Roso},\
  and\ \citenamefont {Plaja}}]{Perez-Hernandez2009}%
  \BibitemOpen
  \bibfield  {author} {\bibinfo {author} {\bibfnamefont {J.~A.}\ \bibnamefont
  {P\'{e}rez-Hern\'{a}ndez}}, \bibinfo {author} {\bibfnamefont
  {L.}~\bibnamefont {Roso}},\ and\ \bibinfo {author} {\bibfnamefont
  {L.}~\bibnamefont {Plaja}},\ }\bibfield  {title} {\bibinfo {title} {Harmonic
  generation beyond the strong-field approximation: the physics behind the
  short-wave-infrared scaling laws},\ }\href
  {https://doi.org/10.1364/OE.17.009891} {\bibfield  {journal} {\bibinfo
  {journal} {Opt. Express}\ }\textbf {\bibinfo {volume} {17}},\ \bibinfo
  {pages} {9891} (\bibinfo {year} {2009})}\BibitemShut {NoStop}%
\bibitem [{\citenamefont {Henneberger}(1968)}]{Henneberger1968}%
  \BibitemOpen
  \bibfield  {author} {\bibinfo {author} {\bibfnamefont {W.~C.}\ \bibnamefont
  {Henneberger}},\ }\bibfield  {title} {\bibinfo {title} {Perturbation method
  for atoms in intense light beams},\ }\href
  {https://doi.org/10.1103/PhysRevLett.21.838} {\bibfield  {journal} {\bibinfo
  {journal} {Phys. Rev. Lett.}\ }\textbf {\bibinfo {volume} {21}},\ \bibinfo
  {pages} {838} (\bibinfo {year} {1968})}\BibitemShut {NoStop}%
\bibitem [{\citenamefont {Reed}\ \emph {et~al.}(1993)\citenamefont {Reed},
  \citenamefont {Burnett},\ and\ \citenamefont {Knight}}]{Reed1993}%
  \BibitemOpen
  \bibfield  {author} {\bibinfo {author} {\bibfnamefont {V.~C.}\ \bibnamefont
  {Reed}}, \bibinfo {author} {\bibfnamefont {K.}~\bibnamefont {Burnett}},\ and\
  \bibinfo {author} {\bibfnamefont {P.~L.}\ \bibnamefont {Knight}},\ }\bibfield
   {title} {\bibinfo {title} {Harmonic generation in the
  $\text{Kramers-Henneberger}$ stabilization regime},\ }\href
  {https://doi.org/10.1103/PhysRevA.47.R34} {\bibfield  {journal} {\bibinfo
  {journal} {Phys. Rev. A}\ }\textbf {\bibinfo {volume} {47}},\ \bibinfo
  {pages} {R34} (\bibinfo {year} {1993})}\BibitemShut {NoStop}%
\bibitem [{\citenamefont {Lakhotia}\ \emph {et~al.}(2020)\citenamefont
  {Lakhotia}, \citenamefont {Kim}, \citenamefont {Zhan}, \citenamefont {Hu},
  \citenamefont {Meng},\ and\ \citenamefont {Goulielmakis}}]{Lakhotia2020}%
  \BibitemOpen
  \bibfield  {author} {\bibinfo {author} {\bibfnamefont {H.}~\bibnamefont
  {Lakhotia}}, \bibinfo {author} {\bibfnamefont {H.~Y.}\ \bibnamefont {Kim}},
  \bibinfo {author} {\bibfnamefont {M.}~\bibnamefont {Zhan}}, \bibinfo {author}
  {\bibfnamefont {S.}~\bibnamefont {Hu}}, \bibinfo {author} {\bibfnamefont
  {S.}~\bibnamefont {Meng}},\ and\ \bibinfo {author} {\bibfnamefont
  {E.}~\bibnamefont {Goulielmakis}},\ }\bibfield  {title} {\bibinfo {title}
  {Laser picoscopy of valence electrons in solids},\ }\href
  {https://doi.org/10.1038/s41586-020-2429-z} {\bibfield  {journal} {\bibinfo
  {journal} {Nature}\ }\textbf {\bibinfo {volume} {583}},\ \bibinfo {pages}
  {55} (\bibinfo {year} {2020})}\BibitemShut {NoStop}%
\bibitem [{\citenamefont {Gaarde}\ \emph {et~al.}(2008)\citenamefont {Gaarde},
  \citenamefont {Tate},\ and\ \citenamefont {Schafer}}]{Gaarde_review}%
  \BibitemOpen
  \bibfield  {author} {\bibinfo {author} {\bibfnamefont {M.~B.}\ \bibnamefont
  {Gaarde}}, \bibinfo {author} {\bibfnamefont {J.~L.}\ \bibnamefont {Tate}},\
  and\ \bibinfo {author} {\bibfnamefont {K.~J.}\ \bibnamefont {Schafer}},\
  }\bibfield  {title} {\bibinfo {title} {Macroscopic aspects of attosecond
  pulse generation},\ }\href {https://doi.org/10.1088/0953-4075/41/13/132001}
  {\bibfield  {journal} {\bibinfo  {journal} {Journal of Physics B: Atomic,
  Molecular and Optical Physics}\ }\textbf {\bibinfo {volume} {41}},\ \bibinfo
  {pages} {132001} (\bibinfo {year} {2008})}\BibitemShut {NoStop}%
\bibitem [{\citenamefont {Baggesen}\ and\ \citenamefont
  {Madsen}(2011)}]{Baggesen2011}%
  \BibitemOpen
  \bibfield  {author} {\bibinfo {author} {\bibfnamefont {J.~C.}\ \bibnamefont
  {Baggesen}}\ and\ \bibinfo {author} {\bibfnamefont {L.~B.}\ \bibnamefont
  {Madsen}},\ }\bibfield  {title} {\bibinfo {title} {On the dipole, velocity
  and acceleration forms in high-order harmonic generation from a single atom
  or molecule},\ }\href {https://doi.org/10.1088/0953-4075/44/11/115601}
  {\bibfield  {journal} {\bibinfo  {journal} {Journal of Physics B: Atomic,
  Molecular and Optical Physics}\ }\textbf {\bibinfo {volume} {44}},\ \bibinfo
  {pages} {115601} (\bibinfo {year} {2011})}\BibitemShut {NoStop}%
\bibitem [{\citenamefont {Madsen}(2021)}]{Madsen2021a}%
  \BibitemOpen
  \bibfield  {author} {\bibinfo {author} {\bibfnamefont {L.~B.}\ \bibnamefont
  {Madsen}},\ }\bibfield  {title} {\bibinfo {title} {Different forms of
  laser--matter interaction operators and expansion in adiabatic states},\
  }\bibfield  {journal} {\bibinfo  {journal} {The European Physical Journal
  Special Topics}\ }\href {https://doi.org/10.1140/epjs/s11734-021-00026-y}
  {10.1140/epjs/s11734-021-00026-y} (\bibinfo {year} {2021})\BibitemShut
  {NoStop}%
\bibitem [{\citenamefont {Kyle}\ and\ \citenamefont
  {McDowell}(1969)}]{Kyle1969}%
  \BibitemOpen
  \bibfield  {author} {\bibinfo {author} {\bibfnamefont {H.~L.}\ \bibnamefont
  {Kyle}}\ and\ \bibinfo {author} {\bibfnamefont {M.~R.~C.}\ \bibnamefont
  {McDowell}},\ }\bibfield  {title} {\bibinfo {title} {On a peaking
  approximation in scattering theory},\ }\href
  {https://doi.org/10.1088/0022-3700/2/1/304} {\bibfield  {journal} {\bibinfo
  {journal} {Journal of Physics B: Atomic and Molecular Physics}\ }\textbf
  {\bibinfo {volume} {2}},\ \bibinfo {pages} {15} (\bibinfo {year}
  {1969})}\BibitemShut {NoStop}%
\end{thebibliography}


%

\end{document}